\documentclass{aa}
\usepackage{graphicx}
\usepackage{epsfig}
\usepackage{graphics}
\usepackage{rotating}
\begin{document}

\title{
Gamma-ray absorption and the origin  of the gamma-ray flare in Cygnus X-1}
\author{Gustavo E. Romero\inst{1,2,}\thanks{Member of CONICET, Argentina}, Maria Victoria del Valle\inst{1,2,}\thanks{Fellow of CONICET, Argentina}  \and Mariana Orellana\inst{2,3} }
  
\institute{Instituto Argentino de Radioastronom\'{\i}a (IAR), CCT La Plata  (CONICET), C.C.5, (1894) Villa Elisa, Buenos Aires, Argentina \and Facultad de Ciencias Astron\'omicas y Geof\'{\i}sicas, Universidad Nacional de La Plata, Paseo del Bosque s/n, 1900, La Plata, Argentina \and 
Departamento de F\'{\i}sica y Astronom\'{\i}a, Universidad de Valpara\'{\i}so, Chile}

\offprints{M. V. del Valle \\ \email{maria@iar-conicet.gov.ar}}

\titlerunning{Gamma-ray flares in Cyg X-1}

\authorrunning{G.E. Romero et al. }

\abstract
{The high-mass microquasar Cyg X-1, the best-established candidate for a stellar-mass black hole in the Galaxy, has been detected in a flaring state at very high energies (VHE), $E>$ 200 GeV, by the Atmospheric Cherenkov Telescope MAGIC. The flare occurred at orbital phase $\phi = 0.91$, where $\phi = 1$ is the configuration with the black hole behind the companion high-mass star, when the absorption of gamma-ray photons by photon-photon annihilation with the stellar field is expected to be  highest.}
{We aim to set up a model for the high-energy emission and absorption in Cyg X-1 that can explain the nature of the observed gamma-ray flare.} 
{We study the gamma-ray opacity due to pair creation along the whole orbit, and for different locations of the emitter. Then we consider a possible mechanism for the production of the VHE emission.} 
{We present detailed calculations of the gamma-ray opacity and infer from these calculations the distance from the black hole where the emitting region was located. We suggest that the flare was the result of a jet-clump interaction where the decay products of inelastic $p-p$ collisions dominate the VHE outcome.} 
{We are able to reproduce the spectrum of Cyg X-1 during the observed flare under reasonable assumptions. The flare may be the first event of jet-cloud interaction ever detected at such high energies.}

 
\maketitle
 
\section{Introduction}
 
Five X-ray binaries have been detected in the very high-energy region of the electromagnetic spectrum, $E \sim$ TeV. Three of them, PSR B1259-63, LS I +61 303 and LS 5039, have been detected at different orbital phases and show variable emission. Four gamma-ray flares were detected by the AGILE satellite from the exceptional X-ray binary Cyg X-3 (Tavani et al. 2009). The Fermi Large Area Telescope (LAT) has also detected a variable high-energy source coinciding with the position of Cyg X-3 (Abdo et al. 2009). The fifth source, Cyg X-1, has been detected only once during a flare episode. This latter detection constitutes the first evidence of very high-energy gamma-ray emission produced in the surroundings of a stellar-mass black hole (BH) in our galaxy (for further discussion see Paredes 2008). 
 
Recently, Albert et al. (2007) reported  the results from  observations of Cyg X-1 at very high energies, $E>200$ GeV, performed with the Major Atmospheric Gamma Imaging Cherenkov (MAGIC) telescope. No persistent emission was detected, but a fast transient episode was. The satellites INTEGRAL and Swift/BAT detected with some delay a related flare at hard X-rays, while only  a statistically poor detection was found in the RXTE/ASM data at soft X-rays. This wavelength-dependent behavior may suggest that different emitting regions were involved. The gamma-ray excess occurred at orbital phase $\phi$ = 0.91. This can help to set constraints on the location of the emission region.  More recently, the flaring nature of Cyg X-1 in gamma rays has been confirmed with the AGILE satellite (Sabatini et al. 2010). This work is devoted to a study of the absorption of high-energy photons in Cyg X-1 and the implications of the resulting constraints.  

The paper is organized as follows: in the next section we describe the main characteristics of the source under study. Section 3 deals with the gamma-ray opacity by pair creation in the stellar radiation field. The production mechanism of the flare emission is then examined in the context of existing models (e.g. Bosch-Ramon, Romero, \& Paredes 2006; Romero et al. 2003). In particular, we explore the physical conditions required by the energy budget and spectrum of the flare event.
In Sect. 4 we present a simple modelization for the non-thermal emission and compare our calculations with the observational results. Finally, in Sect. 5, we present  a brief discussion and the conclusions. 

\section{Cygnus X-1}

The binary system Cyg X-1 is composed by a massive star and a compact object. The X-ray and radio monitoring of the source over the last decades have shown that Cyg X-1 is most of the time in a hard X-ray state and powers collimated jets (e.g. Stirling et al. 2001), which makes it a confirmed high-mass microquasar (HMMQ, Mirabel \& Rodr\'{\i}guez 1999). It is located at a distance of 2.2$\pm$0.2 kpc (Zi\'olkowski 2005). The massive star is an O9.7 Iab of 40$\pm$10 $M_{\odot}$ and the compact object is the best-established candidate for a stellar-mass BH in the Galaxy, with 21$\pm$8 $M_{\odot}$ (Zi\'olkowski 2005). The orbit of the system is circular, with a period of 5.6 days and an inclination between 25$^{\circ}$ and 65$^{\circ}$ (Gies \& Bolton 1986).  At radio wavelengths, a semi-ring surrounds Cyg X-1. This feature is thought to be the result of a strong shock at the location where the jet impacts onto the ambient interstellar medium (Gallo et al. 2005).

Regarding the flare event at VHEs, the observed energy spectrum is well fitted by a relatively soft power law (Albert et al. 2007)
\begin{eqnarray}
	\frac{ {\rm d} N}{{\rm d} A {\rm d} t {\rm d} E } = &&(2.3 \pm 0.6) \times 10^{-12}
 \bigl(\frac{E}{1 {\rm TeV}}\bigr)^{-3.2 \pm 0.6 }\nonumber\\ &&{\rm cm}^{-2} {\rm s}^{-1}{\rm TeV}^{-1}.
\label{A07}
\end{eqnarray}
 
The star provides an intense radiation field that can absorb gamma-rays by pair creation within the binary system. The detection by MAGIC occurred near the superior conjunction, when this opacity to gamma-ray propagation from a region close to the compact object is expected to be maximum. 

The massive star has a strong wind. Considerable observational evidence supports the idea that winds of high-mass stars are clumpy (e.g. Owocki \& Cohen 2006, Moffat 2008). In a HMMQ, some clumps could eventually penetrate into the jet of the system enhancing the non-thermal emission, as proposed by Owocki et al. (2009).

\section{Gamma-ray opacity due to $e^{+}e^{-}$ pair creation in the stellar radiation field}

\subsection{Calculations}

\begin{figure}[!h]
\resizebox{\hsize}{!}{\includegraphics{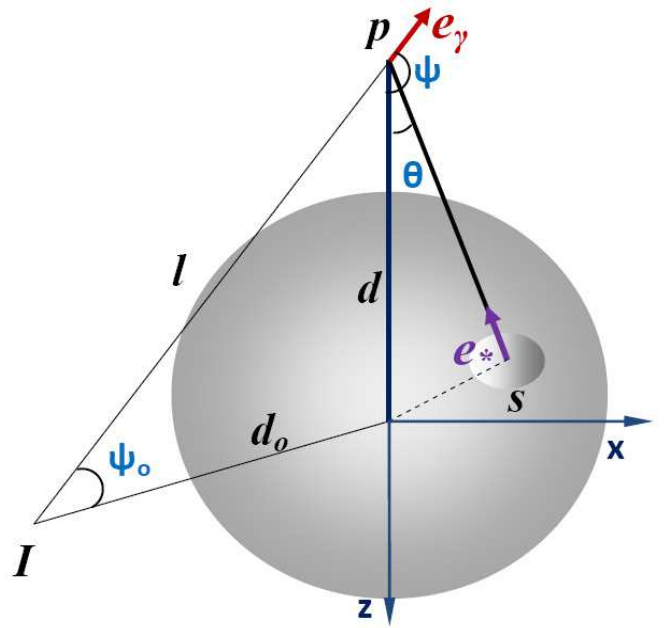}}
\caption{Gamma-ray photon at $P$ travels in the direction given by $\textbf{e}_{\gamma}$. This photon can be absorbed by photons of an energy $\epsilon$ emitted at $S$ in the direction $\textbf{e}_{\star}$. Adapted from Dubus (2006).}
  \label{fig:estrella}
\end{figure}

In a HMMQ the radiation field of the massive star provides soft photons that can annihilate gamma-rays by pair creation: $\gamma+\gamma \rightarrow e^{+} + e^{-}$. We consider the opacity treatment for  gamma-ray absorption in a massive X-ray binary system as in Dubus (2006) and Romero et al. (2007). The differential opacity for a gamma-ray at $P$ traveling in the direction given by $\textbf{e}_{\gamma}$ due to photons of an energy $\epsilon$ emitted at $S$ in the direction $\textbf{e}_{\star}$ is (Fig. \ref{fig:estrella})
\begin{equation}
{\rm d}\tau_{\gamma\gamma} = (1-{\bf e}_{\gamma}{\bf e}_{\star})n_{\epsilon}\sigma_{\gamma\gamma}{\rm d}{\epsilon}{\rm d}{\Omega}{\rm d}l,
\label{dtau}
\end{equation} 
where d$\Omega$ is the solid angle of the surface that emits the photons and $n_{\epsilon}$ is the specific radiation density. 

The cross-section for photon annihilation is (Gould \& Schr\'eder 1967)
\begin{eqnarray}
\sigma_{\gamma\gamma}(\beta) = &&\frac{{\pi}r_{e}^{2}}{2}(1-\beta^{2})\nonumber\\ &&\times\left[2\beta(\beta^{2}-2)+(3-\beta^{4})\ln{\Big(\frac{1+\beta}{1-\beta}\Big)}\right],
\label{cross}
\end{eqnarray}
where $\beta =(1-1/s)^{1/2}$, and $s=E_{\gamma}\epsilon (1-{\bf e}_{\gamma}{\bf e}_{\star})/(m_{e}c^2)^2$. Here, $E_{\gamma}$ and $\epsilon$ are the energies of the gamma-ray and the stellar photon, respectively. This reaction occurs above a minimun energy given by
\begin{equation}
E_{\gamma}\epsilon =\frac{2(m_{e}c^{2})^{2}}{(1-{\bf e}_{\gamma}{\bf e}_{\star})},
\label{umbral}
\end{equation}
where $\textbf{e}_{\gamma}$ is a unitary vector in the direction of the gamma-ray propagation and $\textbf{e}_{\star}$ is also a unitary vector in the direction of the stellar photon propagation. 
The optical depth is a trajectory integral for which the angular dependence has a very significant effect. The absorption is then highly modulated by the orbital motion. It depends also on the target photon field, which is strongly anisotropic along the gamma-ray path . 

Because the massive star completely dominates the spectral distribution of the radiative field at low energies, any other source of radiation for the production of pairs with gamma rays is neglected here. The star has a radius $R_{\star}$, and for simplicity we asume  a blackbody density radiation of a temperature $T_{\star}$:
\begin{equation}
n_{\epsilon} = \frac{2\epsilon^{2}}{h^{3}c^{3}}\frac{1}{({\rm exp}(\epsilon/kT_{\star})-1)}\,\,\,{\rm ph}\,{\rm cm}^{-3}{ \rm erg}^{-1}{\rm sr}^{-1}.
\label{CN}
\end{equation}  

\begin{figure}[!h]
\resizebox{\hsize}{!}{\includegraphics{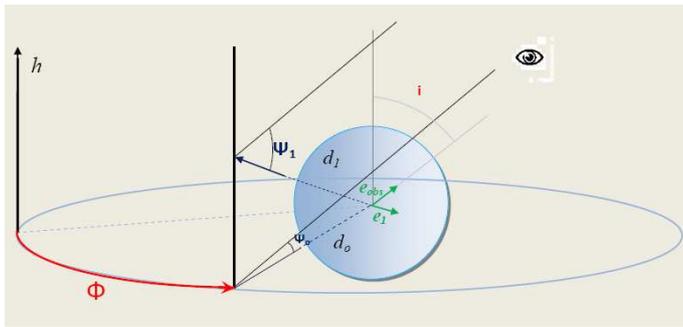}}
\caption{Sketch of the geometry considered for the gamma-ray absorption of a photon that is produced above the compact object.} 
\label{fig:orbita}
\end{figure}

The geometry considered for the gamma-ray absorption is shown in Fig. \ref{fig:orbita}. If emission occurs at a height $h$ above the compact object and perpendicular to the orbital plane, the distance $d$ from the star becomes $d = \sqrt{d_{0}^{2}+h^{2}}$ and the initial angle changes from $\psi_0 = \sin(\theta)\sin(i)$ to $\psi_1$. Note that according to Fig. \ref{fig:orbita} 
\begin{equation}
\cos\psi_{1} = {\bf e}_{1}\cdot{\bf e}_{\rm obs}=\frac{1}{\sqrt{(d_{0}^{2}+h^{2})}}(d_{0}\cos{2\pi \phi}\sin{i}-h\cos{i}).
\end{equation}

\begin{table}
\caption[]{Model parameters. Those related to the absorption are listed first.}
\begin{tabular}{lll}
\hline\noalign{\smallskip}
\multicolumn{2}{l}{Parameter [units]} & values\\[0.001cm]
\hline\noalign{\smallskip}
$R_{\star}$ &Stellar radius [cm] &  1.5$\times10^{12}$  \\[0.001cm]
$T_{\star}$ & Stellar temperature [K] & $3\times10^{4}$  \\[0.001cm] 
$r_{\rm orb}$ &Orbital radius [cm] & $3.4\times10^{12}$  \\[0.001cm]
$\theta$ &Viewing angle & ${\pi}/6$  \\[0.001cm]
\hline\noalign{\smallskip}
$M_{\rm BH}$ &BH mass [M$_{\odot}$]& 20\\[0.001cm]
$h_{0}$ &Jet initial point [R$_{\rm Sch}$] & 50  \\[0.001cm]
$h_{\rm int}$ &Height above compact object [cm] & $10^{13}$  \\[0.001cm]
$\varrho$ &Equipartition parameter & 0.1 \\[0.01cm]
$\Gamma_{\rm jet}$ &Jet bulk Lorentz factor & 1.4\\[0.001cm]
$B$ & Magnetic field [G] & 0.9 \\[0.001cm]
$\eta$ &Acceleration efficiency& 0.1\\[0.001cm]
$L_{\rm jet}$ &Jet kinetic power [erg s$^{-1}$] & $10^{37}$ \\[0.001cm]
$a$ &Hadron-to-lepton energy ratio & 100 \\[0.001cm]
$q_{\rm rel}$ &Jet content of relativistic particles & 5\% \\[0.001cm]
$R_{\rm jet}$ &Jet radius [$h_{\rm int}$] &  0.1  \\[0.001cm]
$e$ &Thickness  of the ``one  zone'' [$h_{\rm int}$] & 0.05  \\[0.001cm]
$\zeta$ &Particle injection index & 2.8\\[0.001cm]
$\dot{M}_{\star}$ &Stellar mass loss rate [M$_{\odot}$yr$^{-1}$] & 3$\times$10$^{-6}$  \\[0.001cm]
$v_{\infty}$ &Terminal wind velocity [cm s$^{-1}$] & $2\times10^{8}$  \\[0.001cm]
\hline\\[0.05cm]
\end{tabular}	
  \label{table}
\end{table}

The parameters adopted for the calculations are shown in Table \ref{table}. 

Under adequate conditions, the absorption, resulting in the creation of energetic pairs, and the Inverse Compton (IC) emission from them, can operate in an effective way to develop electromagnetic cascades which can considerably modify the original gamma-ray spectrum (see e.g. Bednarek 1997 and Orellana et al. 2007 for detail treatments). Electrons with TeV energies in the stellar radiation field may also lead to this situation. 
At TeV energies the rate of electron energy losses in the Klein-Nishina regime is reduced by the diminution of the IC cross-section. The ambient magnetic field must be smaller than a critical value $B_{\rm c}$ for the synchrotron losses  not to overcome the IC ones.
In order to determine if effective electromagnetic cascading can occur within the system it is then necessary to know the magnetic field strength in the gamma-ray propagation region. Such a field is dominated by the stellar magnetic field. Magnetic fields measured in massive stars can reach $\sim 10^3$ G, which is much greater than the critical value $B_{\rm c}$. For close binaries like Cyg X-1 we can expect that $B > B_{\rm c}$ (Bosch-Ramon, Khangulyan, \& Aharonian 2008) over the whole region of gamma-ray production. We here assume that $B>B_c$, and neglect the  effects of electromagnetic cascades, as well as the reprocessing of the absorbed energy by synchrotron radiation. The latter situation was considered by Bosch-Ramon et al. (2008),  who deal with the diffusion of secondary pairs into the system. Zdziarski et al. (2008), on the other hand, do consider that the HE photons iniciate a spatiallty extended pair cascade, but we will comment on this below (Sect. 5). 

\subsection{Results}

\begin{figure}[!ht]
\resizebox{\hsize}{!}{\includegraphics {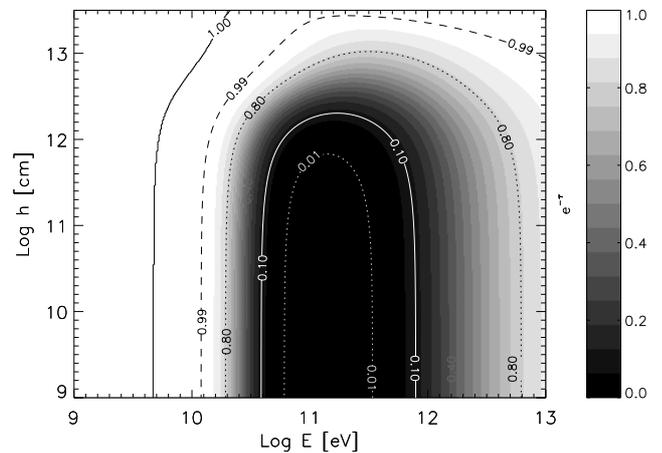}}
\caption{Absorption map as a function of the height $h$ above the compact object and the energy $E$ for orbital phase $\phi = 0.91$.}
\label{fig:Abs}
\end{figure}

In Fig. \ref{fig:Abs} we show a 2D-map of the attenuation coefficient $e^{-\tau}$ as a function of the energy $E$ and the height $h$ above the orbital plane. This absorption map corresponds to the orbital phase $\phi$ = 0.91, when the flare occurred. As can be seen from the figure, the attenuation is high at energies between 10 GeV and 10 TeV, close to the compact object, which makes the absorption problem in the energy range where MAGIC detected the flare very relevant.

\begin{figure}[!ht]
\resizebox{\hsize}{!}{\includegraphics{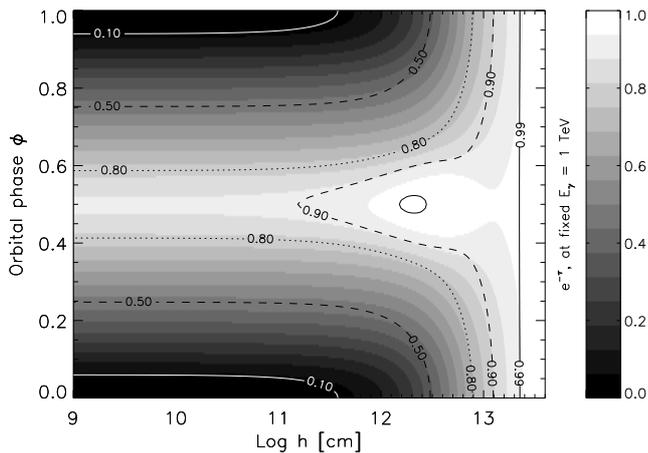}}
\caption{Absorption map as a function of the orbital phase $\phi$ and height $h$ above the compact object for energy $E = 1$ TeV.}
\label{fig:Abs2}
\end{figure}

In Fig. \ref{fig:Abs2} we show  a 2D-map of the attenuation coefficient for $E= 1$ TeV as a function of the orbital phase $\phi$ and the height $h$. It can be seen that the absorption drops strongly as the height above the compact object increases for $h$ $>$ 10$^ {11}$ cm. When $h$ $<$ 10$^{11}$ cm the absorption does not present major changes, due to the distances involved that make the photon density  remain rather constant (i.e. $R_{\star}$ $\sim$ 10$^{12}$ cm and $r_{\rm orb} = 3.4\times10^{12}$ cm; see Fig. \ref{fig:orbita}).
Bosch-Ramon et al. (2008) find out from opacity calculations near the superior conjunction that the TeV emitter in Cyg X-1 should be located at a distance greater than $10^{12}$ cm above the compact object. Our absorption calculations agree with this result. Notice that our results cover a much larger parameter space.

\begin{figure}[!ht]
\resizebox{\hsize}{!}{\includegraphics[angle=270]{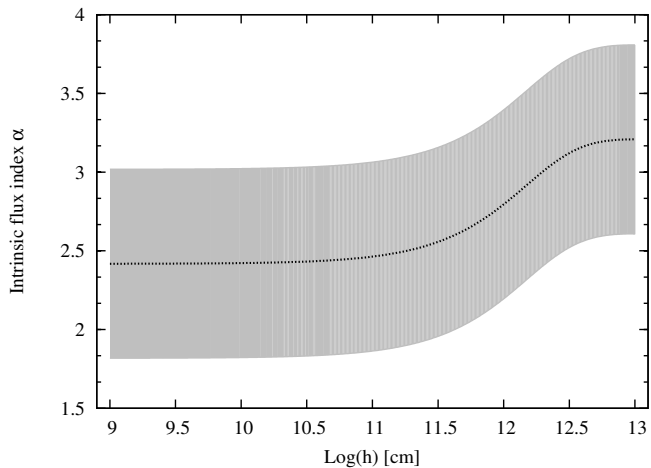}}
\caption{Range within the error bars of the intrinsic flux index as a function of the height $h$.}
\label{index}
\end{figure}

From Albert et al. (2007) the observed flux is a power law (Eq. \ref{A07}), in the energy range between 150 GeV and 3 TeV. Considering that the intrinsic flux from the flare is also a power law $F_{{\rm int}}= {\kappa}E^{-\alpha}$, we can relate both expressions through
\begin{equation}
F_{{\rm obs}}= F_{{\rm int}}e^{-\tau(E)}.
\end{equation}
From the computed  numerical values of $\tau(E)$, using the dependence of $\tau$  on the height $h$, we obtain the intrinsic spectral index $\alpha$ as a function of the latter parameter. Figure \ref{index} shows the  result. 
Note that for an altitude of $h$ $\sim$ 10$^{13}$ cm the de-absorbed and production spectra are essentially the same.

\section{Flare production mechanism}
A hadronic MQ model for Cyg X-1 has been already considered in Orellana et al. (2007) based on ideas advanced by Romero et al. (2003). We here revisit  that scenario with the addition of the interaction between the steady jet and a more dense target: a clump from the stellar wind that  allows through locally generated shocks the reacceleration of the particles that produce VHE emission far from the BH, as in Araudo, Bosch-Ramon, \& Romero (2009). The jet+clump system is assumed to be momentarily in steady state. As observed in the stable configuration of a microquasar in a low-hard X-ray state (e.g. Fender, Belloni, \& Gallo 2004), we assume a continuous jet. The calculations of the  emission are based on the works by Bosch-Ramon et al. (2006) and Romero \& Vila (2008). 

The jet is considered perpendicular to the orbital plane, and  launched at a distance $h_{0}$ above the compact object. 

We consider that farther down the jet the magnetic field reaches values well below equipartition. Following Bosch-Ramon et al. (2006) the magnetic field in the jet reference frame can be calculated as
\begin{equation}
B(h) = \sqrt{{\varrho}8\pi e_{\rm p}}.
\label{subequi}
\end{equation}
In Eq. (\ref{subequi}) $\varrho$ is the equipartition parameter and ${\it e}_{\rm p}$ is the matter energy density. Then, 
\begin{equation}
e_{\rm p} = \frac{\dot{m}_{\rm jet}}{\pi R_{\rm jet}^{2}v_{\rm jet}m_{p}}\langle{E_{\rm p\,k}}\rangle = \frac{\dot{m_{\rm jet}}}{2\pi h^{2}}v_{\rm jet},
\end{equation}
where $v_{\rm jet}$ is the bulk velocity of the outflow, we set $v_{\rm jet}\sim 0.7 c$  (Heinz 2006). The jet radius is $R_{\rm jet} = \chi h$, and $\langle{E_{\rm p\,k}}\rangle$ is the mean kinetic energy of the cold proton,  taken to be the classical kinetic energy with a velocity equal to the expansion velocity of the jet ($v_{\rm exp} = \chi v_{\rm jet}$).

A small fraction of the jet power is transformed into relativistic particles in a ``one-zone'' acceleration region located above the compact object, at the height of the impact with the clump. Here we assumed $h_{\rm int} = 10^{13}$ cm, based on our opacity constraints. 

The kinetic power in the form of relativistic particles is assumed to be proportional to the jet's power, $L_{\rm rel} = q_{\rm rel} L_{\rm jet }$, with $q_{\rm rel} = 0.05$ and $L_{\rm jet}=10^{37}$ erg s$^{-1}$ (Gallo et al. 2005). We considered both hadronic and leptonic content, $L_{\rm rel} = L_{\rm p} + L_{\rm e}$. The ratio of relativistic protons to electrons luminosity ${\it a}$ in the jet is unknown. We adopted $ a = 100$, a similar value to what is observed in the galactic cosmic ray spectrum (e.g. Berezinskii et al. 1990).

The minimum kinetic energy is taken to be on the order of the rest mass energy of the corresponding particles. The maximum energy for the electrons  is obtained equating the cooling rates with the acceleration rate. The acceleration rate by Fermi mechanism, $t_{\rm acc}^{-1} = E^{-1}$d$E$/d$t$, of a particle with energy $E$ in a magnetic field $B$, is given by
\begin{equation}
t_{\rm acc}^{-1} = \frac{\eta e c B}{E},
\end{equation}
with $\eta$  the acceleration efficiency, which is assumed here to be high, $\sim$ 10 \%. 
The maximum energy for protons is restricted by the size of the acceleration region because the particle giroradius $r_{\rm g} = E/eB$ should not exceed $R_{\rm jet}$. 
The energy losses considered for electrons are adiabatic, IC, synchrotron and relativistic Bremsstrahlung, and are calculated in the jet reference frame (RF).\\

For adiabatic losses, the cooling rate is 
\begin{equation}
t_{\rm ad}^{-1} = \frac{2}{3}\frac{v_{\rm jet}}{h_{\rm int}}.
\end{equation}
 
The synchrotron losses rate is
\begin{equation}
t_{\rm synchr}^{-1} = \frac{4}{3}\frac{\sigma_{\rm T}cU_{B}}{m_{\rm e}c^{2}}\biggl(\frac{m_{\rm e}}{m}\biggr)^{3}\frac{E}{mc^{2}},
\end{equation}
where $\sigma_{\rm T}$ is the Thomson cross-section and $U_{B}$ is the magnetic energy density.

The IC loss rate can be calculated from (Blumenthal \& Gould 1970)
 \begin{equation}
t_{\rm IC}^{-1} = \frac{1}{E_{\rm e}}\int_{\epsilon_{\rm min}}^{\epsilon_{\rm max}}
\int_{\epsilon}^{\frac{bE_{\rm e}}{1+b}} (\epsilon_{1} - \epsilon) \frac{{\rm d}N}{{\rm d}t{\rm d}\epsilon_{1}}{\rm d}{\epsilon}_{1} , 
\end{equation}
where $\epsilon$ and  $\epsilon_{1}$ are the incident and scattered photon energies, respectively, and
 
\begin{equation}
\frac{{\rm d}N}{{\rm d}t{\rm d}\epsilon_{1}} = \frac{1}{E_{\rm e}}\frac{2\pi r_{0}^{2}mc^{3}}{\gamma}\frac{n_{\rm ph}(\epsilon){\rm d}\epsilon}{\epsilon}f(q),
\end{equation}
with
\begin{equation}
f(q) = 2q\ln{q}+(1+2q)(1-q)+\frac{1}{2}\frac{(bq)^{2}}{a+bq}(1-q).
\end{equation}
Here $b = 4\epsilon\gamma/mc^{2}$  and $q = \epsilon_{1}/[b(E_{\rm e}-\epsilon_{1})]$.
The seed photon field is provided by the companion star, considered as a blackbody at $T_{\star}$ and is transformed to the jet reference frame (Dermer \& Schlickeiser 2002)
\begin{equation}
n'_{\epsilon',\Omega'} = \frac{n_{\epsilon, \Omega}}{\Gamma_{\rm jet}^{2}(1+\beta\mu)^2},
\end{equation}
where $\Gamma_{\rm jet}$ is the jet Lorentz factor, $\Omega$ represents the photon direction, $\mu = \cos{\Theta}$, and $\Theta$ is the angle between the photon direction and the jet axis (the quantities with primes are in the jet RF). We considered the ``head on'' approximation, in which $\mu = -1$. 
 
The relativistic Bremsstrahlung losses for a complete ionized plasma were computed according to (Berezinskii et al. 1990)
\begin{equation}
t_{\rm Br}^{-1} = 4nZ^{2}r_{e}^{2}\alpha c \biggl[\ln{\frac{2E_{\rm e}}{m_{\rm e}c^{2}}}-\frac{1}{3}\biggr],
\end{equation}
where $n=n_{\rm wind}/\Gamma$ is the density of target ions expressed in the jet RF. For external Bremsstrahlung the target-ion density is that of the stellar wind ions (the wind is considered as a completely ionized plasma). At a height $h$

\begin{equation}
n_{\rm wind}(h)= \frac{\dot{M_{\star}}}{4\pi v_{\infty }m_{p}(h^{2}+a^{2})}\left(1-\frac{R_{\star}}{\sqrt{h^{2}+{r_{\rm orb}}^{2}}}\right)^{-1},
\end{equation} 
where  $v_{\infty}$ is the terminal velocity of the wind (e.g. Romero et al. 2003 and references therein). In order to take into account the mixing between the jet and the background wind material  in a phenomenological way, we introduced a penetration factor $f_{\rm p} = 0.3$ (Romero, Christiansen, \& Orellana 2005). The clump is considered as a condensation of the wind with a density of $\sim$ 10$^{14}$ cm$^{-3}$. Notice that particle rejection at the jet-wind boundary is considered only for the background wind. The clump is assumed to fully penetrate the jet (see Araudo et al. 2009 for details). For internal Bremsstrahlung the target-ion density is the proton density of the jet, directly derived from $\dot{m}_{\rm jet} = L_{\rm jet}/\Gamma c^{2}$.

Relativistic protons lose energy through adiabatic expansion, synchrotron radiation, and by losses produced by hadronic interactions. The energy loss rate produced by proton-proton interactions is
\begin{equation}
t_{pp}^{-1} = n_{p}c\sigma_{pp}K_{pp},
\end{equation}
where $n_{p}$ is the density of target protons and $K_{pp}$ the inelasticity ($\sim$ 0.5). The cross-section can be approximated (Kelner, Aharonian \& Bugayov 2006) by
\begin{equation}
\sigma_{pp} = (34.3+1.88L+0.25L^{2})\biggl[1-\biggl(\frac{E_{\rm th}}{E_{p}}\biggr)^{4}\biggr]^{2} {\mbox mb},
\label{sigma_pp}
\end{equation}
where $L = \ln(E_{p}/1{\rm TeV})$. Photomeson production is not considered because the stellar photons do not have enough energy to reach the threshold energy of this process.
\begin{figure}[!ht]
\resizebox{\hsize}{!}{\includegraphics[angle=270]{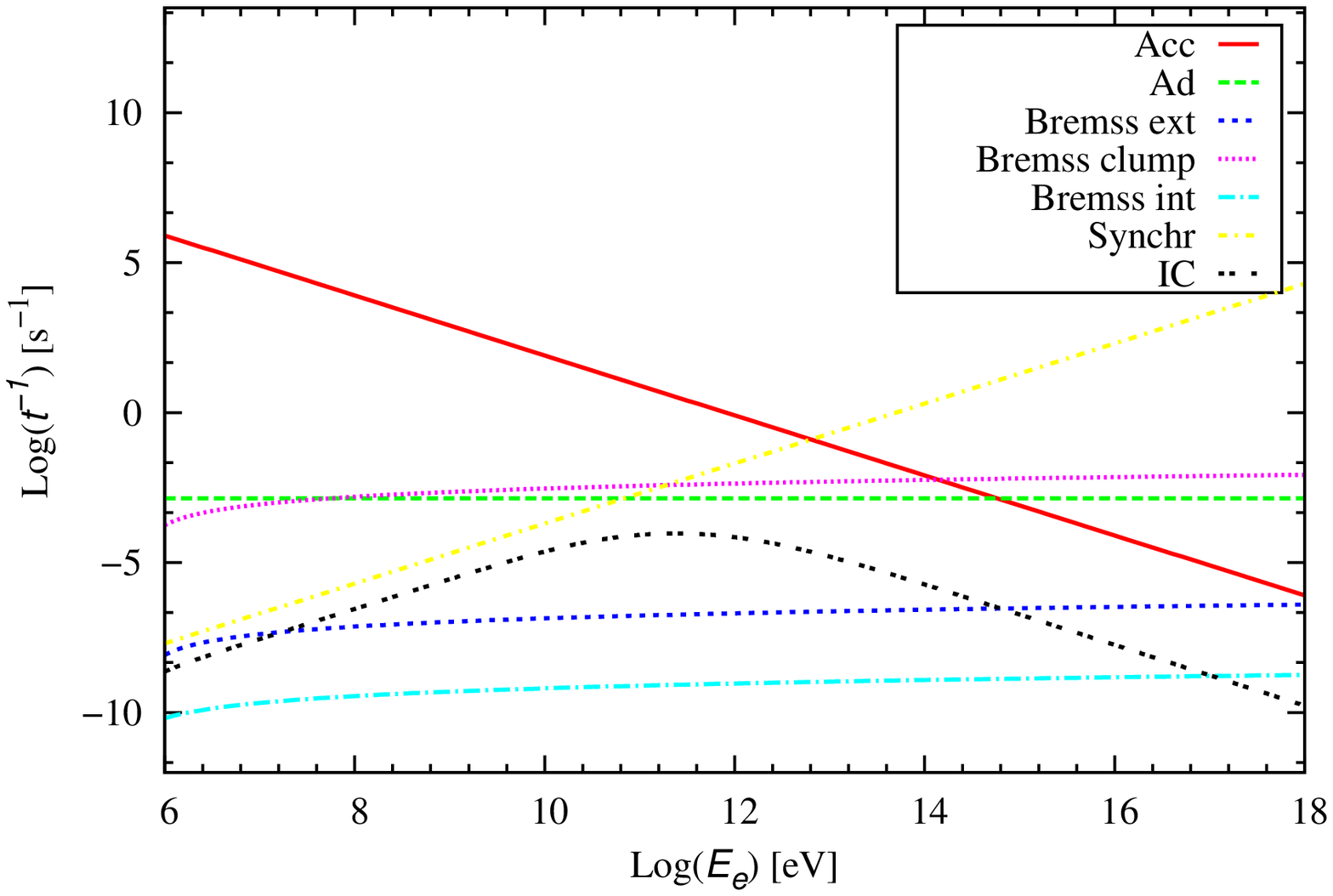}}
\resizebox{\hsize}{!}{\includegraphics[angle=270]{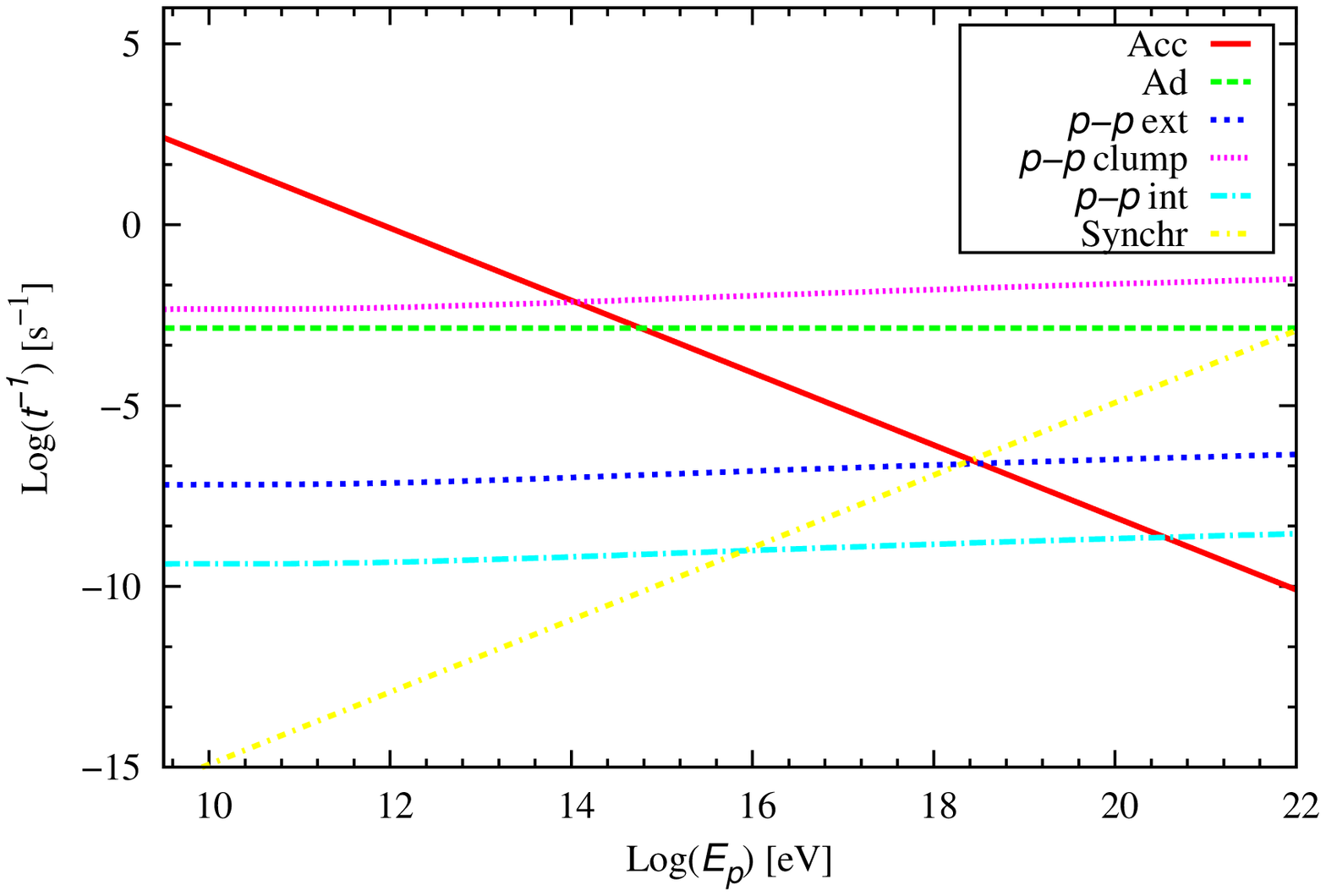}}
  \caption{Acceleration and cooling rates at $h_{\rm int} = 10^{13}$cm in the jet for primary electrons and protons.}
  \label{fig:cool}
\end{figure}

In Fig. \ref{fig:cool} we show the rates of cooling and energy gain for electrons and protons in the acceleration region, which is considered to be the bow-shock between the jet and the clump. The electrons reach TeV energies while the protons can attain energies $\sim$ 10$^{2}$ TeV. 

In the one zone approximation the steady state particle distributions $N(E)$ result from the transport equation (Ginzburg \& Syrovatskii 1964)
\begin{equation}
 \frac{\partial}{\partial E}\biggl[\frac{{\rm d}E}{{\rm d}t}{\bigg\arrowvert}_{\rm loss}N(E)\biggr]+\frac{N(E)}{t_{\rm esc}} = Q(E),
\end{equation}
where $t_{\rm esc} \sim h_{\rm int}/v_{\rm jet}$. 

The exact analytical solution of the equation is
\begin{eqnarray}
N(E)= &\biggl\arrowvert \frac{{\rm d}E}{{\rm d}t}\biggl{\arrowvert}_{\rm loss}^{-1}\int_{E}^{E^{\rm max}}{\rm d}E'& Q(E')\nonumber\\ 
& &\times{\exp}\biggl(-\frac{\tau(E,E')}{t_{\rm esc}}\biggr),
\end{eqnarray}
with
\begin{equation}
\tau(E,E')= \int_{E}^{E'} {\rm d}E'' \biggl\arrowvert \frac{{\rm d}E''}{{\rm d}t}\biggr{\arrowvert}_{\rm loss}^{-1}. 
\end{equation}
The particle injection function, $Q(E)$, is assumed to be a power-law in the energy of the particles, 
\begin{equation}
Q(E)=Q_{0}\,E^{-\zeta}.
\end{equation}
This distribution is expected to be the result of diffusive particle acceleration by the reverse shock. The index $\zeta$ is taken as 2.8 for both types of particles, in accordance with the obtained results shown in Fig. \ref{index}.
The normalization constant $Q_{0}$ for each type of particle is derived from $L_{{\rm e},p}$ as 
\begin{equation}
L_{{\rm e},p} = V \int_{E_{{\rm e},p}^{\rm min}}^{E_{{\rm e},p}^{\rm max}}
{\rm d}E_{{\rm e},p}E_{{\rm e},p} Q_ {{\rm e},p}(E_{{\rm e},p}),
\end{equation}
where $V$ is the co-moving one-zone volume. 

\subsection{Radiative processes}

\begin{figure*}[]
\centering
\resizebox{\hsize}{!}{\includegraphics[angle=270]{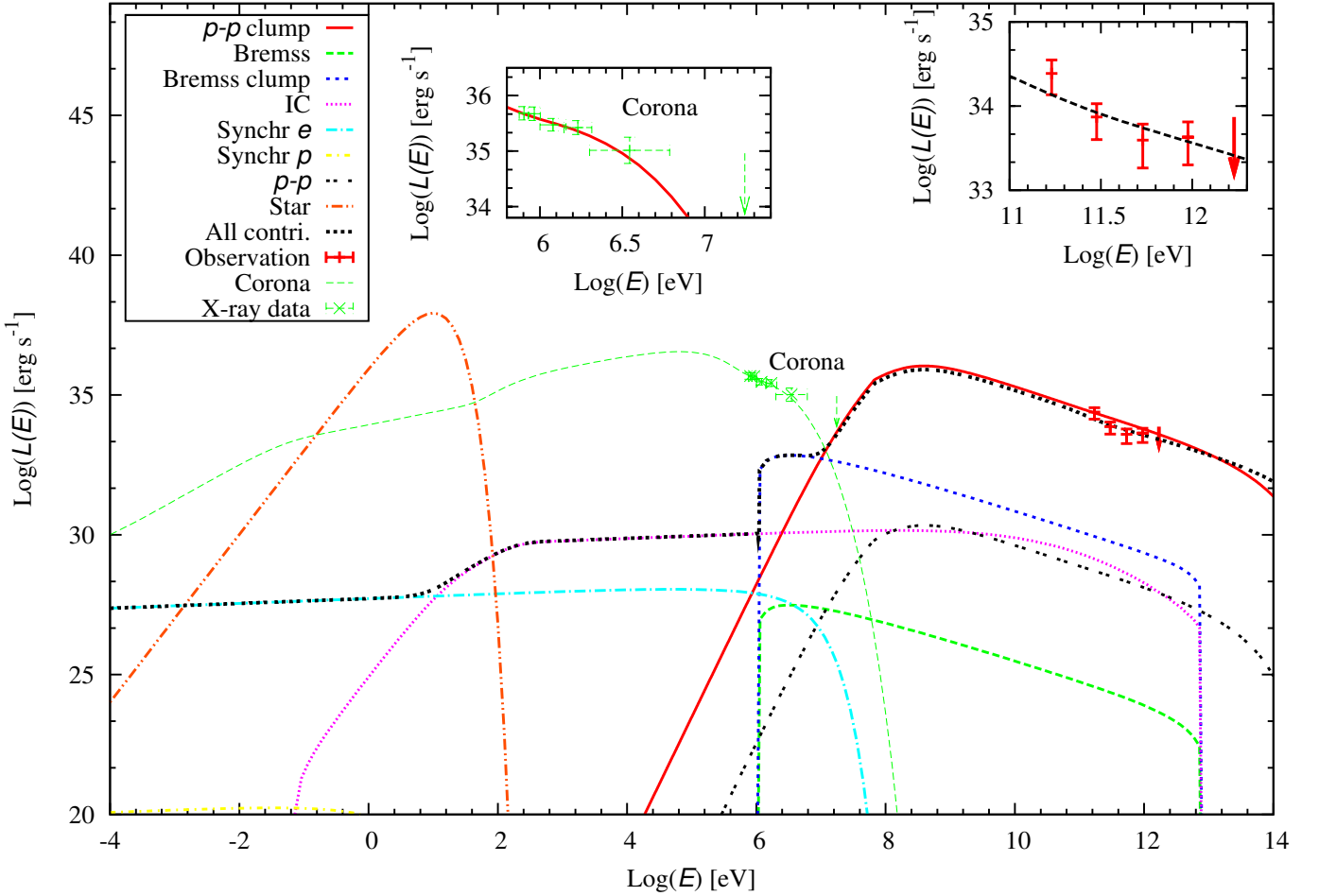}}
\caption{Computed SED and the MAGIC observational data from Cyg X-1 (Albert et al. 2007). A two-temperatures corona with a non-thermal component is presented as well. The data are from McConnell et al. (2000). The similar data from Malzac et al. (2008) can be easily fitted (see Romero, Vieyro, \& Vila 2010).}
\label{SED}
\end{figure*}

We consider synchrotron emission from both electrons and protons, inverse Compton  emission from electron interactions  with the stellar photon field, internal and external relativistic Bremsstrahlung, and inelastic collisions between relativistic protons in the jet and the cold material that forms the jet, plus with the matter of the clump and the background wind. We checked that the emission produced by secondary particles is negligible, as well as the synchrotron self-Compton (SSC).\\

The synchrotron emission was computed with  the approximation
\begin{eqnarray}
L_ {\gamma}(E_ {\gamma}) = && E_ {\gamma} V \frac{\sqrt{3}e^{3}B}{hmc^{2}}\int_{E_{\rm min}}^{E_ {\rm max}} 
{\rm d}E N(E)\frac{E_ {\gamma}}{E_ {\rm c}}1.85\nonumber\\&&\times\biggl(\frac{E_ {\gamma}}{E_ {\rm c}}\biggr)^{1/3}{\exp}\biggl(\frac{E_ {\gamma}}{E_ {\rm c}}\biggr),
\label{Syn}
\end{eqnarray}
where
\begin{equation}
E_ {\rm c} = \frac{3}{4\pi}\frac{ehB}{mc}\biggl(\frac{E}{mc^{2}}\biggr)^{2},
\end{equation}
and the usual meaning for the constants $c$, $h$, $e$.\\

The IC emission by the electron population was calculated as
\begin{eqnarray}
L_{\rm IC}(E_ {\gamma}) = && E_{\gamma}^{2} V \int_{E_{\rm min}}^{E_{\rm max}}{\rm d}E_{\rm e}N_{\rm e}(E_{\rm e})\nonumber\\&&\times\int_{\epsilon_{\rm min}}^{\epsilon_{\rm max}}{\rm d}\epsilon
P_ {\rm IC}(E_{\rm e},E_{\gamma},\epsilon),
\label{IC}
\end{eqnarray}
where the spectrum of scattered photons is
\begin{equation}
P_{\rm IC}(E_{\rm e},E_{\gamma},\epsilon) = \frac{3\sigma_{\rm T}c(m_{\rm e}c^{2})^{2}}{4E_{\rm e}^{2}}
\frac{n_{\rm ph}(\epsilon)}{\epsilon} F(q),
\end{equation}
with
\begin{equation}
F(q) = 2q\ln{q}+(1+2q)(1-q)+\frac{1}{2}(1-q)\frac{(q\Omega)^{2}}{(1+\Omega q)},
\end{equation}
and $\Omega = (4\epsilon E_{\rm e})/((m_{\rm e}c^{2})^{2})$ , $ q = (E_{\gamma})/[({\Omega}E_{\rm e}(1-E_{\gamma}/E_{\rm e}))]$.\\

The relativistic Bremsstrahlung contribution is given by
\begin{equation}
L_{\gamma}(E_{\gamma})= E_{\gamma}V \int_{E_{\gamma}}^{\infty} n \sigma_{\rm B}(E_{\rm e},E_{\gamma})
\frac{c}{4\pi}N_{\rm e}(E_{\rm e}){\rm d}E_{\rm e},
\end{equation}
where 
\begin{equation}
\sigma_{\rm B}(E_{\rm e}, E_{\gamma}) = \frac{4\alpha r_{\rm 0}^{2}}{E_{\gamma}}\phi(E_{\rm e}, E_{\gamma}),
\end{equation}
and 
\begin{eqnarray}
\phi(E_{\rm e}, E_{\gamma}) = &&[1+(1-E_{\gamma}/E_{\rm e})^{2}-2/3(1-E_{\gamma}/E_{\rm e})]\nonumber\\ 
&&\times\biggl\{\ln{\frac{2E_{\rm e}(E_{\rm e}-E_{\gamma})}{m_{\rm e}c^{2}E_{\gamma}}} -\frac{1}{2}\biggr\}.
\end{eqnarray}

All  luminosities were calculated in the jet co-moving RF. 
Photon energies in both frames are related by the Doppler factor $D$ as 
\begin{equation}
E_{\gamma} = DE'_{\gamma},
\end{equation}
where 
\begin{equation}
D = \frac{1}{{\Gamma}_{\rm jet}(1-{\beta}_{\rm jet}\cos{\theta_p})} . 
\end{equation}
The luminosity in the observer frame is given by (e.g. Lind \& Blandford 1985)
\begin{equation}
L_{\gamma}(E_{\gamma}) = D^{2}L'_{\gamma}(E'_{\gamma}).
\end{equation}

In order to  compute the gamma-ray emission produced by neutral pion decay we note that the $p-p$ cross-section parametrization ( Eq. (\ref{sigma_pp})) is given in the laboratory RF. Then, we convert the flux of relativistic protons to the laboratory frame: 
\begin{equation}
J(E_{p},\theta_{p}) = \frac{A{\Gamma}^{-(\alpha-1)}\biggl(E_{p} - \beta\cos{\theta_{p}}\sqrt{E_{p }^{2}-m_{p}^{2}c^{4}}\biggr)^{-\alpha}}
{4\pi\biggl[ \sin{\theta_{p}^{2}}+{\Gamma}^{2}\biggl(\cos{\theta_{p}}-\frac{{\beta}E_{p}}{\sqrt{E_{p}^{2}-m_{p}^{2}c^{4}}}\biggr)^{2}\biggr]^{1/2}
},
\end{equation}
where $A$ is a normalization constant. The flux of protons, which is isotropic in the jet RF, is beamed in the lab RF, as indicated by the dependence on the viewing angle $\theta_{p}$.

The gamma-ray luminosity, for $E_{p} <$ 0.1 TeV,  can be obtained straightforwardly as 
\begin{equation}
L_{\gamma}(E_{\gamma}) = V E_{\gamma}^{2} 2 \int_{E_{\min}}^{\infty} \frac{q_{\pi}(E_{\pi})}{\sqrt{E_{\pi}^{2}-m_{\pi}^{2}c^{4}}} {\rm d} E_{\pi},
\end{equation}
with $E_{\rm min}= E_{\gamma}+m_{\pi}c^{4}/4E_{\pi}$.
In the formalism of the $\delta$-functional  (Aharonian \& Atoyan 2000) the $\pi^{0}$-emissivity is given by
\begin{equation}
q_{\pi}(E_{\pi}) = \frac{n_{\rm p}}{\kappa_{\pi}}\sigma_{pp} \biggl(m_{p}c^{2}+E_{\pi}/\kappa_{\pi} \biggr)J_{p}\biggl(m_{ p}c^{2}+E_{\pi}/\kappa_{\pi}\biggr)
\end{equation}
with $\kappa_{\pi} \sim $ 0.17 (Gaisser 1990). For $E_{ p} <$ 0.1 TeV down to the threshold, a slightly modified version of $\delta$-functional approximation is needed, using the replacement
\begin{equation}
\delta(E_{\pi}-{\kappa}_{\pi}E_{\rm kin}) \rightarrow \tilde{n}\delta(E_{\pi}-{\kappa}_{\pi}E_{\rm kin}).
\end{equation}
Here $\tilde{n}$ is the total number of $\pi^{0}$ created per $p-p$ collision.

The gamma-ray luminosity in the range 0.1 TeV$\le E_{p} \le 10^{5}$ TeV can be obtained from (Kelner et al. 2006)
\begin{eqnarray}
L_{\gamma}(E_{\gamma}) = && n_{p} E_{\gamma}^{2} V \int_{E_{\gamma}}^{\infty} \sigma_{\rm inel}(E_{p})
J_{p}(E_{p})\nonumber\\&&\times F_{\gamma}\biggl(\frac{E_{\gamma}}{E_{p}},E_{p}\biggr) \frac{{\rm d}E_{p}}{E_{p}},
\end{eqnarray}
with $F_{\gamma}\biggl(\frac{E_{\gamma}}{E_{p}},E_{p}\biggr)$ a function of $E_{\gamma}$ and $E_{p}$. 
For further details on radiative processes see Vila \&  Aharonian (2009) and references therein. 

In order to reproduce the observed  spectral energy distribution (SED), the density ratio between the clump and the wind at the base is $\sim$ 4.6$\times$10$^{4}$, i.e. $n_{\rm c} \sim 3.3\times 10^{14}$ cm$^{-3}$. 


Figure \ref{SED} shows the computed SED. We have included the thermal emission by the star, which largely dominates at optical energies. At X-rays, the components of the emission by the accretion disk and a corona should be added to our results. These components in the low-hard state have luminosities $\sim 10^{37}$ erg s$^{-1}$ and extend up to $\sim$ 150 keV (see Romero et al. 2002), in  a way that they completely dominate over the non-thermal radiation. The emission from the corona and a non-thermal tail (McConnell et al. 2000, Malzac et al. 2008) are also shown. The model for this emission is from Romero, Vieyro, \& Vila (2010) and is presented in detail elsewhere. Here we show only the results relevant to Cyg X-1. 

\subsection{Internal absorption}
Internal photon-photon annihilation within the  region of gamma-ray production can result in strong attenuation of the radiation (Aharonian et al. 2008, Romero \& Vila 2008). The opacity is again an integral of Eq. (\ref{dtau}), but now considering  the locally produced photons with density $n_{\rm ph}(\epsilon)$. We can use the symmetry in one of the angles to write
\begin{equation}
\tau (E_{\gamma}) =  \frac{1}{2}\int_{l} \int_{\epsilon_{\rm th}}^{\epsilon_{\rm max}} \int_{-1}^{u_{\rm max}} (1-u) \, \sigma_{\gamma\gamma}(\beta)
n_{\rm ph}(\epsilon) {\rm d}u {\rm d}\epsilon {\rm d}l .
\end{equation}
Here, $u = \cos \vartheta$, $\vartheta$ is the angle between the momenta of the colliding photons, $l$ is the photon path, and the cross-section $\sigma_{\gamma\gamma}(\beta)$ is given by Eq. (\ref{cross}). The absorbing photon fields are those generated within the jets (i.e. those calculated in the former section). At energies $E_\gamma$ $\ga$ 10$^{15}$ eV, the dominant absorbing field is the synchrotron radiation from electrons. In the local approximation of Ghisellini et al. (1985), 
\begin{equation}
n_{\rm synchr} \approx \frac{\epsilon_{\rm synchr}}{\epsilon}\frac{r}{c}, 
\end{equation}
where $\epsilon_{\rm synchr}$ is the synchrotron power per unit volume per unit energy: $ \epsilon_{\rm synchr} = L_{\gamma}/(\epsilon^{2} V)$, with $L_{\gamma}$ from Eq.(\ref{Syn}). 

The geometry considered requires $r = R_{\rm jet}$ and $0 \le l \le R_{\rm jet}$.
We find that $\tau(E_{\gamma})$ is completely negligible (at the level of $\tau$ $\sim$ 10$^{-6}$),  implying that the attenuation coefficient is $\sim$ 1.

\section{Discussion}

The VHE transient emission of Cyg X-1 occurred when the BH was behind the star with respect to the observer. Because of the high absorption in the flare detection energy range, the emission close to the BH is not enough to explain the observations, unless the photons travel far away from the star, initiating a spatially extended pair cascade as considered by Zdziarski et al. (2008). This requires a fine tunned magnetic field, which allows the instantaneous isotropization of the electrons, but does not overcome their IC radiative losses. A more realistic/accurate calculation of the electromagnetic cascade propagation is then desirable. Such simulations (following the electron trajectories) will be available in a future work as an application of the code developed by Pellizza et al. (2009).
Previous 1D cascade simulations (Orellana et al. 2007) are  consistent with a strong absorption and steep spectrum at TeV energies. The results by Bosch-Ramon et al (2008) have shown that if the cascades are suppressed by effects of the magnetic field, the synchrotron emission of the secondary pairs peaks at lower energies ($\sim$ GeV).

Romero, Kaufman-Bernad\'o, \& Mirabel (2002) have suggested that Cyg X-1 could go through occasional microblazar phases and have estimated that the luminosity in the observer RF can be up to one order of magnitude higher than the luminosity in the jet RF. Even taking this into
account, a flare triggered at the base of the jet is undetectable due to absorption at phase 0.91. A remaining option could be  a very short episode with a highly increased acretion/ejection rate, but this is speculative given the lack of evidence at  lower energies supporting the hypothesis.

Under the geometry considered here (a jet perpendicular to the orbital plane, which has an inclination of $\sim$ 30 deg), the high-energy emission should have occurred at a large distance above the compact object where the absorbing photon field is diluted. In order to quantify the radiative outcome in this scenario we have considered the interaction of relativistic particles accelerated in a narrow region of the jet with the target particles of a dense clump of the wind.

The flare timescale is related to the permanence of the clump inside the jet. For a spherical clump with a radius
$R_c$ smaller than the jet radius $R_{\rm jet}$ $\sim 10^{12}$ cm we can make a zerolth order estimation of the time that it takes the clump to cross the jet: $t_c$. The clump velocity is the wind velocity, which at this height is simply $v_{\infty}$:
\begin{equation}
t_c\simeq 2R_{\rm jet}/v_{\infty}\sim 10^4 \, {\mbox s}.
\end{equation}
The flaring episode had a timescale shorter than one day and a rising time of about one hour, which is on the same order as the $t_c$ estimated. 

The simple model presented here for the broadband spectrum of Cygnus X-1  reproduces fairly well the observed  SED by MAGIC during the flare using a set of parameters that agrees with reasonable  values for this source. Interactions between the clumpy winds of massive stars with the relativistic jets in HMMQ are expected to be  produce flaring episodes at high and very high energies, and may be detectable by the new high-energy detectors, like Fermi, MAGIC II, and VERITAS.

\begin{acknowledgements}
We thank Florencia Vieyro and Gabriela Vila for help on several aspects of this work. We also thank an anonymous referee for valuable comments. This work was partially supported by grant ANPCyT (PICT 2007-00848, BID 1728/OC-AR). G.E.R. acknowledges support from the Ministerio de Educaci\'on y Ciencia (Spain) under grant AYA 2007-68034-C03-01 FEDER funds. 
\end{acknowledgements}

\end{document}